\documentclass[12pt]{article}
\usepackage{amsmath,amssymb,pgf,pgfarrows,pgfnodes,subfig,float,appendix, hyperref}
\usepackage[margin=0.7in]{geometry}

\title{{\rm\footnotesize \qquad \qquad \qquad \qquad \qquad \ \qquad \qquad \qquad \ \ \ \ \ \                  UTTG-31-13\ TCC-025-13     RUNHETC-2013-17     
SCIPP 13/11}\vskip.5in     Holographic Space-time and Newton's Law}
\author{Tom Banks\\
Department of Physics and SCIPP\\
University of California, Santa Cruz, CA 95064\\
{\it and}\\
Department of Physics and NHETC\\
Rutgers University, Piscataway, NJ 08854\\
E-mail: \href{mailto:banks@scipp.ucsc.edu}{banks@scipp.ucsc.edu}
\\
\\
W. Fischler\\
Department of Physics and Texas Cosmology Center\\
University of Texas, Austin, TX 78712\\
E-mail: \href{mailto:fischler@physics.utexas.edu}{fischler@physics.utexas.edu}}

\begin{document}
\maketitle

\begin{abstract}
We derive Newton's Law from the formalism of Holographic Space-Time (HST).  More precisely, we show that for a large class of Hamiltonians of the type proposed previously for the HST description of a geodesic in Minkowski space, the eikonal for scattering of two massless particles at large impact parameter scales as expected with the impact parameter and the energies of the particles in the center of mass (CM) frame. We also discuss the criteria for black hole production in this collision, and find an estimate, purely within the HST framework, for the impact parameter at which it sets in, which coincides with the estimate based on general relativity\cite{penroseetal}.
\end{abstract}

\section{Introduction}

In a recent paper\cite{holounruh} the authors proposed a class of time dependent Hamiltonians, which are candidates to describe the physics of time-like geodesics in the Holographic Space-Time (HST)\cite{HST} formalism for $4$ dimensional Minkowski space. The Hamiltonians are all based on a definition of ``particle" in the HST formalism, which we review below, and all of them give rise to particle S-matrices with a conserved energy and a crude form of locality/cluster decomposition. The axioms of HST require more than this.  Consistency between the evolution along different Minkowski geodesics, combined with the observation that the overlap of the particle sub-factor of each trajectory's Hilbert space, has complete asymptotic overlap with the particle sub-factor of any other trajectory,  implies that the S-matrix be the same for all geodesics, or simply that the particle Hilbert space of any trajectory, carry a unitary representation of the super-Poincare group, which commutes with the S-matrix.  The Poincare transformed states for one trajectory, correspond to the states of another.

We have not yet implemented these constraints, but experience from string theory suggests that they will be very strong, and considerably reduce the ambiguity in the Hamiltonians we have proposed so far.  It is therefore somewhat surprising that, in this paper, we will be able to show that the leading contribution to two body scattering at large impact parameter, has the same parametric dependence on center of mass (c.m.) energy and impact parameter as one computes from the Born graph with graviton exchange in the t-channel.
That is to say, all of these Hamiltonians give rise to Newton's law of gravity.
Given the particle content of our model, results of Mandelstam and Weinberg\cite{mandelwein} show that Newton's law is an inevitable consequence of the Lorentz invariance of the correct Hamiltonian within our class, but our
derivation works for {\it all} of the Hamiltonians in our class, despite the fact that most of them are not Lorentz invariant. Although we will not go into the derivation here, it seems likely that this leading behavior will exponentiate in the standard fashion to given the eikonal approximation to large impact parameter scattering.  We believe that the generality of our result is a consequence of the fact that much of the physics of gravity is encoded in the relation between geometry and entropy discovered by Bekenstein and Hawking\cite{bhjfsbpv}.  This law is the basis of HST.

HST describes a space-time quantum mechanically in terms of an infinite number of quantum systems, each of which represents the view of space-time from the vantage point of a single time-like trajectory.  The Hamiltonian that propagates in proper time along a trajectory is time dependent, and has the form
$$ H (n, {\bf x})  = H_{in} (n, {\bf x}) + H_{out} (n, {\bf x}) .$$ ${\bf x}$ is a label for different trajectories.  For the present paper, which deals with scattering in Minkowski space, we can take it to be a cubic spatial lattice.  The real spatial geometry is defined by causality conditions, and those will be defined such that the locus of all points on the lattice that are a fixed number, $D$, of steps apart, is a sphere of radius $D$ in Planck units.

At the time labelled $\pm n$, $H_{in} (n, {\bf x}$, which propagates the system one Planck step into the future/past of $\pm n$, acts on a Hilbert space of entropy (entropy = ${\rm ln}$ of the dimension) $n^{d-2} L$, which is a unitary representation of a super-algebra, whose Fermionic generators are labelled by the sections of the spinor bundle over the holographic screen, with an eigenvalue cutoff for the Dirac operator\cite{tbjk}.  For Minkowski space, ${\cal M}^d$ the screen is a $d - 2$ sphere times a compact manifold ${\cal K}$.  The spinor generators and their algebra can be displayed as (see below)
    $$\bigl[ \psi_a^{(m_1 \ldots m_{d-2})} [P] , \psi^{\dagger\ b}_{(n_1 \ldots n_{d-2})} [Q] \bigr]_+ = \delta_a^b \delta^{m_1}_{n_1} \ldots \delta^{m_{d-2}}_{n_{d-2}} Z[P,Q]$$ where $a$ labels the components of the minimal $SO(d-2)$ spinor, and the other indices are totally anti-symmetric and take values from $1$ to $n$.  $P$ labels the cutoff spinor bundle on ${\cal K}$.  The even generators of the algebra $Z[P,Q]$ are even functions of the fermions, and form a closed super-algebra, whose structure defines the fuzzy geometry of ${\cal K}$.  For compactifications from $11$ dimensions to $d$ on tori of minimal size, $P$ takes on only a few values, and the $Z[P,Q]$ are just numbers.  For fixed values of the
    non-compact indices, $L$ is the entropy of the minimal representation of this super-algebra.

For very large time, the dimension of the full Hilbert space goes to infinity $n \rightarrow N \rightarrow \infty$.  The Hamiltonian $H_{out} (n)$  acts on the tensor complement of the Hilbert space generated by the variables above, in the Hilbert space generated by the larger set of variables where the labels go from $1$ to $N$, instead of $n$.  Particle states are defined by an asymptotic constraint.   For $d = 4$  the constraint is
$$\psi_{mn} [P] | particle \rangle = 0, $$ for $K N + Q$ variables, where $1 \ll K, Q \ll N$.  Using large $N$ scaling arguments, we showed in \cite{holounruh} that
 $K$ is an asymptotically conserved quantity, proportional to the energy flowing out through the asymptotic holographic screen for null infinity.  In $d$ dimensions, the corresponding constraint is on $N^{d - 3} E$ variables, which can be thought of as follows. The anti-symmetric tensor character of the $\psi$ variables allows us to think of them as living on a $d-2$ cube of side $N$. Particle variables correspond to variables in a smaller hypercube of side $K_i$, one such cube per particle, strung out along the main diagonal of the large hypercube.  We will see below that the particle energies are proportional to $K_i^{d-3}$ .   The last hyper-cube along the diagonal has size $N - \sum K_i$.
 We call these the {\it active horizon DOF}.  The constrained variables correspond to off main block diagonal elements of the large hypercube and their number is   $\sum K_i^{d-3} N^{d-3} $ plus lower order polynomials in $N$.  An argument similar to that of \cite{holounruh} shows that, with the Hamiltonians we use, $\sum K_i^{d-3}$ is conserved, and the change in the lower order polynomial corresponds to a rearrangement of the number of individual particles and their energies.

 In previous papers, we have argued that the spinor variables can be used to construct the Fock space of supersymmetric particles in the large $N$ limit.  One way to understand this is that the spinor bundle on the holographic screen of
 a causal diamond in Minkowski space is a natural fuzzification of the super-BMS algebra\cite{awadagibbons} of null infinity. The super-BMS algebra describes flows of spin and momentum through null infinity.  $\delta$-function localized generators of it correspond to individual particles.   The Bose and Fermi statistics follow from this description of particles as excitations of fields on null infinity, in the same way that they follow in ordinary field theory: particles are just localized excitations of fields and configurations with two particles exchanged correspond to the same field configuration.  The connection between spin and statistics is enforced by the graded nature of the super-BMS algebra.  The algebra has a natural $(-1)^F$ gauge symmetry and the odd generators carry half integer spin.

 The Hamiltonian that we write for a given time-like geodesic has the form
 $$H_{in} (n) = \sum P_0^i  + \frac{1}{n^{d-3}} {\rm tr}\ f (M) ,$$ where
$$M^m_n = \psi_a^{(m, m_1 \ldots m_{d-3})} \psi^{\dagger\ a}_{(n, n_1 \ldots n_{d-3})}  ,$$ is an $n \times n$ matrix.  $f(M)$ is a polynomial of fixed, $n$ independent, order, greater than or equal to $7$.  Its coefficients become $n$ independent for large $n$, and are chosen so that, if the Hamiltonian were time independent, it would be a fast scrambler\cite{sekinosusskind}.

This is the form of the Hamiltonian in a sector with a fixed number of asymptotic particles (we are implicitly assuming $n$ is large), and the $P_0^i$ are the energies of these particles in the rest frame of the chosen geodesic.  We will be concerned in this paper, with processes in which two particles collide at very large impact parameter $ b \gg E^{d-3}$, in Planck units. The free part of the Hamiltonian will, to a good approximation, describe the propagation of these two massless particles in the same way they would be described in relativistic single particle quantum mechanics.  In particular, the transverse spread of the wave function can be controlled by going to a boosted frame.  It makes sense to think of the particles as following fixed light-like geodesics with a fixed impact parameter.  We will thus be describing the eikonal regime of particle scattering, though in this paper we will only calculate the Born approximation to the eikonal amplitude.

\section{Newton's Law}

Consider a pair of massless particles, colliding at very large impact parameter, $b$.  
The momentum space wave functions of the particles are localized in a small solid angle on the holographic screen in the remote past.  For large impact parameter the centers of these angularly localized packets are barely changed by the scattering.  If we consider linear combinations of different longitudinal momenta which are localized, then the particle can be thought of as traveling along a geodesic in Minkowski space.   The two geodesics have a minimal transverse separation equal to the impact parameter at some instant of proper time. We call this the point of impact.

In HST, we will treat this scattering process by studying two time-like geodesics, at relative rest, which pass through the two massless trajectories at the time when they are separated by the impacted parameter. We synchronize the time of the two time-like geodesics, and consider the point of impact to be $t=0$. All causal diamonds to be considered are centered around this point in time.

Consider, along one of the geodesics, which we call trajectory $1$, a causal diamond of size $ n \ll b$ much smaller than the impact parameter.  The time dependent Hamiltonian, $H_{in}^1 (n)$ describing evolution inside this diamond, does not act on the variables representing the second massless particle.  It describes only free propagation of the particle whose trajectory intersects this geodesic at the point of impact. 

The consistency conditions of HST for causal diamonds of size $N \gg b$ can be satisfied by saying that the Hamiltonian $H_{out}^1 (n)$ for trajectory $1$ contains a subsystem, whose Hilbert space state is identical to that on which the Hamiltonian $H_{in}^2 (n)$ for the second trajectory acts.   However, when $n \geq b$, the small causal diamonds of the two trajectories overlap, and each of the Hamiltonians $H_{in}^{1,2} (n \geq b) $ must act on Hilbert spaces which contain subsystems describing both of the massless particles.

It is at this point that the particles can interact, and the interaction occurs through the multi-linear terms $g_k {\rm tr}\ M^k$ in the Hamiltonian.  As we have pointed out in the introduction, the effect of these terms is bounded by $\frac{1}{n}$.  We conclude that for this whole class of Hamiltonians, the leading eikonal phase scales at most like $\frac{1}{b}$.  We will find that many terms fall off more rapidly with $b$.

Write the matrix $M$ in the following block form

$$\begin{pmatrix}  M_1 & M_{12} & M_{1H} \\ M_{21} & M_2 & M_{2H} \\ M_{H1} & M_{H2} & M_H \end{pmatrix} .$$ Here $M_i$ are $K_i \times K_i$ matrices representing the massless particle DOF, and $H$ is an $(n - K) \times (n - K)$ matrix representing horizon DOF in the causal diamond of size $n$.  $K = K_1 + K_2$ is the maximal\footnote{The actual particle energy depends on the state of the variables that form the matrix $M$, and can range from $K_i / n$ to $K_i$ in Planck units.} total particle energy in Planck units and we will take $K \ll n$ for the moment.  This is the statement that the particle energy in the causal diamond has a Schwarzschild radius much less than $n$.

In order to get a matrix whose trace is order $n$, but which also correlates the two particle entries $M_{1,2}$, we need a polynomial of at least order $7$, and we get an eikonal operator of the form
                $${\cal E} = \frac{1}{n^2} {\rm tr}\ [ M_{H1} M_1 M_{1H} M_H M_{H2} M_2 M_{2H}] . $$  This operator has of order $n K_1 K_2$ terms.  This will be true for all higher order polynomials which correlate $M_{1,2}$ and have $o(n)$ terms.  We can never get a larger number of terms.
                
 Other terms, which correlate $M_{1,2}$ only via $M_{12}$ and $M_{21}$, are suppressed by a factor $1/n$.  Recalling that $ n \geq b$, we see that these terms do not contribute to the leading large impact parameter behavior of the eikonal phase.  Terms which do not correlate $M_{1,2}$, contribute to the analog of disconnected propagator diagrams in field theory.     Terms which do not contain either $M_{1,2}$, describe only the evolution of the horizon states.
 
 Recall that the incoming state is a tensor product of a state of the particle DOF and the horizon DOF.  We are describing elastic eikonal scattering here, so this identification does not change during the course of evolution.  Processes which involve momentum transfer or change in the number of particles are more complicated, and involve mixing up of the particle and horizon DOF in small enough causal diamonds.  In those processes, they do not retain their individual identities.  In the eikonal approximation, valid for large impact parameters, the initial tensor factorization of the Hilbert space is preserved.   We now {\it assume} that the state of the horizon is generic.  That is, that the initial density matrix of the horizon DOF is maximally uncertain.  Since the Hamiltonian is a fast scrambler, this should be true of the final state as well.  Thus, we evaluate an effective interaction between the particles by taking the expectation value of the operator ${\cal E}$ in the maximally uncertain density matrix of horizon states, obtaining an eikonal operator ${\cal E} (M_1 , M_2) $ which depends only on the single particle DOF.  We have\footnote{We note that ${\cal E}$ is a Hamiltonian operator, and not the eikonal phase itself.  The phase comes from evolution with this operator over the time in the causal diamond, which is $\sim b$.  Thus, the eikonal phase itself has another power of $b$ in the numerator.}
  $$ {\cal E} (M_1, M_2 ) = \frac{1}{b} {\rm tr}\ M_1 G_{12} M_2 G_{21} .$$
  The numerical matrices $G_{ij}$, which are not functions of the particle variables, come from a sum of averages over the uncertain density matrix, of all the terms appearing in the Hamiltonian $H_{in} (n)$, whose polynomial order is $\geq 7$.  It's clear that this eikonal has roughly the form expected from single graviton exchange.  The order of magnitude of the interaction is proportional to the product of the maximal energies of the two massless particles, divided by the impact parameter.  We'll see in the next section that this generalizes to $d$ space-time dimensions, with $b \rightarrow b^{d-3}$.   It is also clear that the precise form of the $G_{ij}$ depends on the coefficients $g_k$ in $H_{in}$ . 
  
We have argued that the consistency conditions of HST, imply Lorentz and spatial translation invariance of the scattering matrix.  The $g_k$ must be chosen to enforce these conditions, and ancient arguments of Mandelstam and Weinberg\cite{mandelwein} tell us that these conditions, along with the unitarity and approximate locality that is built into HST, determine the leading long distance behavior of the interaction uniquely.   We have seen that our formalism will definitely produce sub-leading terms in the long distance expansion.  For example, terms in the Hamiltonian of the form
$$\frac{g_4 }{n^2}{\rm tr}\ M_1 M_{12} M_2 M_{21} ,$$ as well as higher order terms in perturbation theory in the $g_k$, will contribute at sub-leading order.
If we have imposed Poincare invariance, according to the rules of HST, then these terms will be attributable to higher derivative terms in a generally covariant effective action.   What we have shown in this paper is that the scaling behavior, both with respect to impact parameter, and particle energy, follows generally from the entire class of Hamiltonians we have proposed.  

Our result also shows that the eikonal approximation to the S-matrix of these systems satisfies (the massless version of), cluster decomposition:  two scattering processes, separated by a distance much larger than their individual
  impact parameters, will have an eikonal operator that is simply the sum of those for the individual processes. Furthermore, if we have three particles, with one much further away from the other pair than their separation, the analog of the analysis we've just done, shows that the third particle interacts with the sum of the energy operators of the other two.
  
  \subsection{Black Hole Formation}
  
  We now relax the condition that $K_i \ll n$.  There is, in HST, an obvious, kinematic, limit $K = \sum K_i < n$.  This follows simply from the counting of rows in an $n \times n$ matrix.  What is remarkable is that, translated into particle physics/GR terms, this simple kinematics is simply Penrose's criterion\cite{penroseetal} for black hole formation in particle collisions. Recalling that we are assuming $n \sim b$, the impact parameter, this bound on $K$ is the statement that the impact parameter is of order the Schwarzschild radius of the c.m. energy.  
  
   In terms of the mechanics of HST, when the $K_i$ are of order $n$, there are no horizon DOF to integrate out.  The idea that the particles are independent weakly interacting systems is manifestly untrue in the Hamiltonian of the model.   {\it Black hole formation thus corresponds to a situation when most of the available DOF in a causal diamond are in mutual interaction.}   The natural time scale for equilibration of incoming particles with the full set of DOF is $n$, and if the Hamiltonian is a fast scrambler\cite{sekinosusskind} then all traces of individual particle identity disappear in a time of order $n {\rm ln} n$.
   
 Speaking in broader terms than a two body collision between particles, we can make the following observation.  Whenever any collection of incoming particles is completely contained in a causal diamond whose radius $n$ is smaller than the Schwarzschild radius of the c.m. energy, the particles lose their identity and we can say that ``a black hole is formed".  If $n$ is not too large, then at a slightly later time, on the future boundary of a somewhat larger diamond, we will again be able to interpret the configuration as a set of outgoing particles.  In an effective theory, on length scales $>> n$, but still microscopic, this process will look like a multi-particle interaction: a local amplitude for some number of incoming particles , to turn into some other multi-particle state.  In diamonds much larger than $n$ there will still be many constraints on the $\psi_i^A$ variables and an approximate separation between particles and horizon states.
 To use picturesque language: {\it In HST, effective particle interactions, all come from the creation and evaporation of Planck scale black holes.}  We caution the reader that there is NO reason to expect that one can use the Hawking formulae to estimate the properties of these microscopic black holes.
  
   \section{Higher Dimensions}
   
   The discussion of the previous section seems almost too good to be true.  Penrose's sophisticated GR argument for the threshold of black hole formation falls out as a simple consequence of the kinematics of matrices.  In this section we show that this miracle is not an accident.  There is a simple generalization of our results to arbitrary dimensions\footnote{The restriction to $d \leq 11$ comes from the imposition of Lorentz invariance.   Our models become kinematically SUSic in any number of dimensions, but this means that for $ d > 11$ they contain massless higher spin particles.  There are no consistent Lorentz invariant interacting low energy field theories for such particles, and so, by the arguments of Mandelstam and Weinberg\cite{mandelwein}, there are no Lorentz invariant unitary S-matrices.}, which is based on the kinematics of the cut-off spinor bundle on the $d-2$ sphere.  
   
   For $d-2$ even the eigenvalues of the Dirac operator on the sphere are
$\pm (M + \frac{d - 2}{ 2})$
where $M$ is a non-negative integer. The degeneracy of this eigen-space is
$$D_{d} (M)= 2^{\frac{d - 2}{2}}\frac{(d - 3 + M)!} {M!(d-3)! }.$$
For $d-2$ odd we have eigenvalues $±(M + d - 2)$,
with degeneracy
  $$D_{d} (M)= 2^{\frac{d - 2}{2}}\frac{(d - 3 + M)!} {M!(d-3)! }.$$ 
  The only difference between the two cases, is the prefactor, which is the dimension, $d_S$  of the spinor representation.  
  
  The cut-off spinor bundle is the sum of these eigen-spaces up to some maximal value of $M$, which we will call $n$.  It is proportional to the radius of the sphere 
  in Planck units.  The dimension of this cutoff space is, up to a multiplicity counting the DOF of the internal dimensions,
  $$S = d_S  \frac{(d - 2 + n)!} {n!(d-2)! }.$$ Apart from the spinor dimension, this is the counting of totally anti-symmetric tensors of rank $d-2$ in $n$ dimensions.
  The $d-2$ dimensional surface of the sphere is apparent in the dimensionality of the spinor representation.   We therefore introduce the notation
  $$\psi^a_{(m_1 \ldots m_{d-2})} ,$$ and 
  $$\psi_a^{\dagger\ (m_1 \ldots m_{d-2})} ,$$
  for the pixel variables in $d$ space-time dimensions.  The reality of the spinor representation jumps with dimension, and in some dimensions the conjugate variable is the same as the original one.  We will ignore these subtleties, since they do not affect the large $n$ counting that gives rise to Newton's law and Penrose's criterion.
  
  Now introduce the matrices   
  $$M_m^n = \psi^a_{(m m_1 \ldots m_{d-3})}  \psi_a^{\dagger\ (n m_1 \ldots m_{d-3})}.  $$  These can be viewed as a gluing of two $d-2$ cubes along $d - 3$ of their faces, defining a link in the orthogonal dimension.  Our ansatz for the HST Hamiltonian in a causal diamond of size $n$ in $d$ Minkowski dimensions
 is
 $$ H = \sum P_0^i  + \frac{1}{n^{2 (d - 3)}} {\rm tr}\ \sum g_k M^k .$$   
 
 There is a large and growing literature on large $n$ tensor models\cite{tensor models} and models with interactions of this type have been studied quite extensively.  In the appendix, we give our own derivation of the fact, well known to the cognoscenti, that with a single factor of $n^{d - 3}$ in the denominator, the 
 interaction would be of order $1$ in the large $n$ limit.  The essential point is that, although $M$ is an $n \times n$ matrix, each of its matrix elements  is a sum of $n^{d - 3}$ terms.  
 
 An important consequence of this counting is that it changes the relationship between the size of a block $K_i$ and the energy.   The bilinears in $\psi$, which define the free Hamiltonian $P_0^i $ have $d - 3$ indices contracted together, and are linear combinations of the matrix elements $M_m^n$.  Thus $E_i \sim K_i^{d - 3}$.   We will also see the same sort of expression appearing in the formula for Newton's law.
 
 The other difference between $d$ and $4$ dimensions is that the constraint of block diagonal $M$, forces the $\psi$ variables to vanish except in blocks along the main diagonal of the $d-2$ cube.  This follows most easily from our picture of constructing the $M$ matrices by gluing along $d- 3$ faces.   We take the vanishing of $\psi$ variables off the main block diagonal to be our definition of particle states in $d$ dimensions.   It follows immediately that $\sum K_i = n$ for the blocks along the main diagonal.  We only have particles if this is achieved by 
 $$\sum K_i = K \ll n$$ for all but one of the blocks.  The last block, whose size is $n - K \gg 1$ represents the horizon degrees of freedom.   Putting this together with the connection between energy and block size, Penrose's criterion 
 for the threshold impact parameter at which black hole formation occurs, becomes $$ b_{th}^{d-3} = E,$$ in $d$ dimensions.  The remarkable ease with which this result falls out of the HST formalism persists!
 
 Our analysis of the dominant interactions at large impact parameter generalizes immediately because it is formulated in terms of the matrix $M$.  Only the estimate of the size of the matrix elements of this matrix changes.  The dominant interaction comes from monomials of order $7$ and higher, and when averaged over the incoming state of the horizon variables it leads to an eikonal interaction
 
 $${\cal E} = \frac{1}{b^{2(d-3)}} {\rm tr} M_1 G_{12} M_2 G_{21} $$,  which is of order
 $$ \frac{E_1 E_2}{b^{(d - 3)}} .$$  Newton's law holds in arbitrary dimension.

\section{Conclusions}

Holographic space-time has degrees of freedom describing Fock spaces of supersymmetric massless particles in Minkowski space of arbitrary dimensions.
We have defined a class of Hamiltonians which describe scattering of these particles, with a conserved energy.  Spatial translation invariance, and Lorentz invariance of the scattering amplitudes are not guaranteed, and are unlikely to be achieved by most of the Hamiltonians in our class.

Nonetheless, we have shown that all of the amplitudes defined by our procedure have a leading behavior at large impact parameter, which obeys Newton's law in a parametric fashion.  The scaling with impact parameter and particle energy is precisely that expected from single graviton exchange.  The actual mechanism of scattering has to do with a large class of ``horizon" states, which are not contained in effective field theory, although they have asymptotically vanishing energy.  

All of our models also contain a mechanism for black hole formation, which sets in parametrically at the impact parameter equal to the Schwarzschild radius of the c.m. energy.  This is Penrose's criterion for black hole formation in collisions.
Black hole formation, in the HST context sets in when the DOF describing particles saturate the covariant entropy bound in small enough causal diamonds.  The distinction between particle and horizon DOF only makes sense asymptotically, in large causal diamonds, where of order $n$ of the DOF are frozen.  

If the mixing of particle and horizon DOF occurs in a small enough diamond, then Hawking's law of evaporation does not hold, the time scale for the existence of the ``micro black hole" is of order the size of the diamond, and on larger scales, the process just looks like a quasi-local multi-particle interaction.
Once one has reached a macroscopic sized black hole, a truly meta-stable excitation is formed and effective field theory descriptions of the process miss the horizon DOF.

\section{Appendix}

Consider an action for rectangular $K \times n$ matrix variables $T_K^a$
$$ S = T_K^a P_a^b T^K_b  + \alpha {\rm tr}\ V(M) ,$$ where $M$ is the $n\times n$ matrix $$M^a_b = T_K^a T^K_b .$$ The variables may be complex, in which case the transpose $T^K_b$ is also complex conjugated. They may be either bosonic or fermionic .   We take $K \gg n$.  

Write $$e^{ \alpha {\rm tr}\ V(M)} = \int\ [dX]  e^{ {\rm tr}\ (XM) + \beta {\rm tr}\ \tilde{V}(X) } ,$$ where the integral is taken over a contour in the $X$ variables, such that it converges.  Then we can do the gaussian $T$ integral and write the partition function of the model as

$$Z = \int\ [dX] e^{ {\rm tr}\ [ \sigma K {\rm ln}\ (P + X) + \beta \tilde{V} (X)]} ,$$
where $\sigma = \pm 1, \pm \frac{1}{2} $, depending on whether the variables are fermionic, bosonic, complex, or real.  Clearly, if we take $\beta$ of order $K$, we can do the $X$ integral in a systematic steepest descents expansion in powers of $\frac{1}{K}$.  Similar expansions are easily derived for correlation functions.   If we undo the $T$ integral, we can still do a systematic expansion of the $X$ integral, from which we conclude that $\alpha \sim \frac{1}{K}$.  

The application of this general result to tensor models of the form we studied in the text, simply takes $K \sim n^{d-3}$.   In those models, the interaction term carries an extra power of $\frac{1}{K}$ so that the 't Hooft coupling is small.  From the HST point of view, this is the adiabatic switching off of the interaction between particles and horizon DOF, which allows for the existence of a unitary particle scattering matrix.

\begin{center} {\bf Acknowledgments }
\end{center}
 The work of T.B. was supported in part by the Department of Energy.   The work of W.F. was supported {\it in part} by the TCC and by the NSF under Grant PHY-0969020.

\end{document}